\documentclass[a4paper,12pt]{article}

\usepackage{amsmath,amssymb}

\usepackage{pdflscape}

\usepackage{amsmath}
\usepackage[mathscr]{eucal}

\usepackage{lmodern}
\usepackage{extarrows}
\usepackage{rotating}
\usepackage{MnSymbol}
\usepackage{caption}
\usepackage{float}

\usepackage{xcolor}

\begin{document}

\begin{center}

\section*{Quantum vortex in a fluid flow: \\ negative effective mass and a novel mechanism for turbulence formation.  }

\vspace{3mm}

{S.V. Talalov}

\vspace{3mm}

{\small Institute of Digital Technology, Togliatti State University, \\ 14 Belorusskaya str.,
 Tolyatti, Samara region, 445020 Russia.\\
svt\_19@mail.ru}

\begin{abstract}
We explore the movement of a thin, circular quantum vortex filament within an infinite cylindrical pipe.
  The fluid surrounding the vortex ring moves through the pipe at a non-zero velocity denoted by $v$. Our study examines the energy spectrum $E = E(p)$, where $p$ represents the total momentum of a vortex ring.
		We have demonstrated that the function $E(p)$ significantly depends on the velocity $v$. The discovered spectrum $E(p)$ reveals the existence of states with both negative and extremely large effective masses.  We also explored the hypothesis regarding the existence of coupled vortex pairs possessing finite summary effective masses. Every pair consists of vortices that possess both positive and negative masses, with the magnitude of these masses being unrestricted.
		In our model, the criterion  for  the appearance of these states is based on  comparing two numbers. The first is seen as a quantum counterpart to the Reynolds number, while the second represents its critical value for a flow with a single vortex. 
We also explore how this studied  effect  might contribute to the emergence of quantum turbulence.
This study discusses a method for determining the critical Reynolds number in quantum turbulence, using the proposed model as a framework. Here, we use a new quantization technique for classical closed vortex filaments developed by the author earlier \cite{Tal22_1,Tal_PhScr24}.
\end {abstract}


{\bf keywords:} ~   quantum vortices,  ~negative effective masses, quantum turbulence.
 
\end{center}


 {\bf PACS numbers:}   47.10.Df    ~~47.32.C

\vspace{5mm}

 \subsection*{Introduction}

 The description of a  turbulent media as a complex  vortex tangle \cite{Feyn_2} requires  a good understanding of the different features of a vortex motion. The deep understanding of vortex structure and their dynamics is crucial, especially in the field of quantum turbulence.  Indeed, as opposed to classical hydrodynamics where the vortex motion is a well-understood and well-describable phenomenon, in a quantum context the situation becomes more ambiguous. Generally speaking, this statement applies to any complex dynamical system, because {\it''\dots methods of quantization are all of the nature of practical rules, whose application depends on consideration of simplicity''} \cite{Dirac}. As it seems, this  Dirac's note has  not lost their relevance in the present day. 
For example, the quantization of a harmonic oscillator using action-angle variables results in completely unexpected outcomes \cite{Kast}. It is clear that a vortex filament is essentially a more complex system than  an harmonic oscillator. In this context, Dirac's ''simplicity principle'' appears less clear in vortex quantization than in some other dynamical systems.

The conception of a quantum vortex, as some kind of topological defect, is probably the most popular approach at the moment (see, for example, \cite{Sonin,Nemir}).
We will not delve into a comprehensive review of the papers on this subject in this study. 
Nevertheless,  we  mention here  the problem of determining the vortex energy. The work \cite{Tsu08} outlines several approaches to this issue.
In a certain sense, the corresponding energy spectrum can be considered complementary to the phonon-roton Landau spectrum \cite{MarPar}. It seems natural, because the excitations in the form of vortex rings can occur in superfluids \cite{RfRf}.
{~ As one of the reasons for this excitation, the passage of charged particles can be considered \cite{RfRf2}.}

Nevertheless, the intricate nature of quantum turbulence drives the quest for new methods. 
In a series of recent studies, the author proposed a new approach to the quantization of small-perturbed vortex rings (see, for example, the paper \cite{Tal22_1} and others) with small, but non-zero core radius ${\sf a}$.
The key and novel  elements of the proposed method include the following points.

First, we consider quantum vortex filaments to be classical systems that are quantized according to the rules of  quantum mechanics. Unlike conventional methods, quantum vortex filaments are not defined as topological defects.  The main difference between our results and the standard ones is that the spectrum of the quantized circulation $\Gamma$ is richer than the well-known spectrum   $\Gamma =\hbar n/\mu_H$ ($n=1,2,...$)   with equidistant levels. The arguments that justify this nontrivial result were discussed in detail in the  previous author's papers. To avoid repetition, we will not be repeating these arguments here.

Second, we avoid the conventional method of defining the thin vortex energy due to the ambiguity of the regularization procedure when the vortex core ${\sf a}$ tends to zero.
Instead, we apply a group-theoretical approach to define  this physical value.
 To fulfill this program, we define a set of independent Hamiltonian variables so that a vortex loop can be described as a particle with internal degrees of freedom. In this context, we define the centrally-extended Galilean group, $\widetilde {\mathcal{G}}_3$, as the space-time symmetry group of the theory. 
As well-known, the Cazimir functions of the 
Lee algebra of the  group $\widetilde{\mathcal G}_3$ with a central charge $\mu_0$ are:  
		 \[ {\hat C}_1 = \mu_0\, {\hat I}\,,\quad 
  {\hat C}_2 = \left[{\hat M}_i  - \sum_{k,j=x,y,z}\epsilon_{ijk}{\hat P}_j {\hat B}_k\right]^{\,\,2} 
  \quad {\hat C}_3 = \hat H -  \frac{1}{2\mu_0}\sum_{i=x,y,z}{\hat P}_i^{\,2}\,.\]                        
    In these formulas, the operators ${\hat M}_i$, $\hat H$, ${\hat P}_i$, and ${\hat B}_i$ (where $i = x, y, z$) represent rotations, time translations, space translations, and Galilean boosts, respectively. The symbol ${\hat I}$ denotes the identity operator.
The Cazimir function ${\hat C}_3$ is often associated with the ''internal energy of the particle''. The central charge $\mu_0$ is  interpreted as a some (conditional) mass.

\begin{center}
		 \includegraphics[width=2.0in]{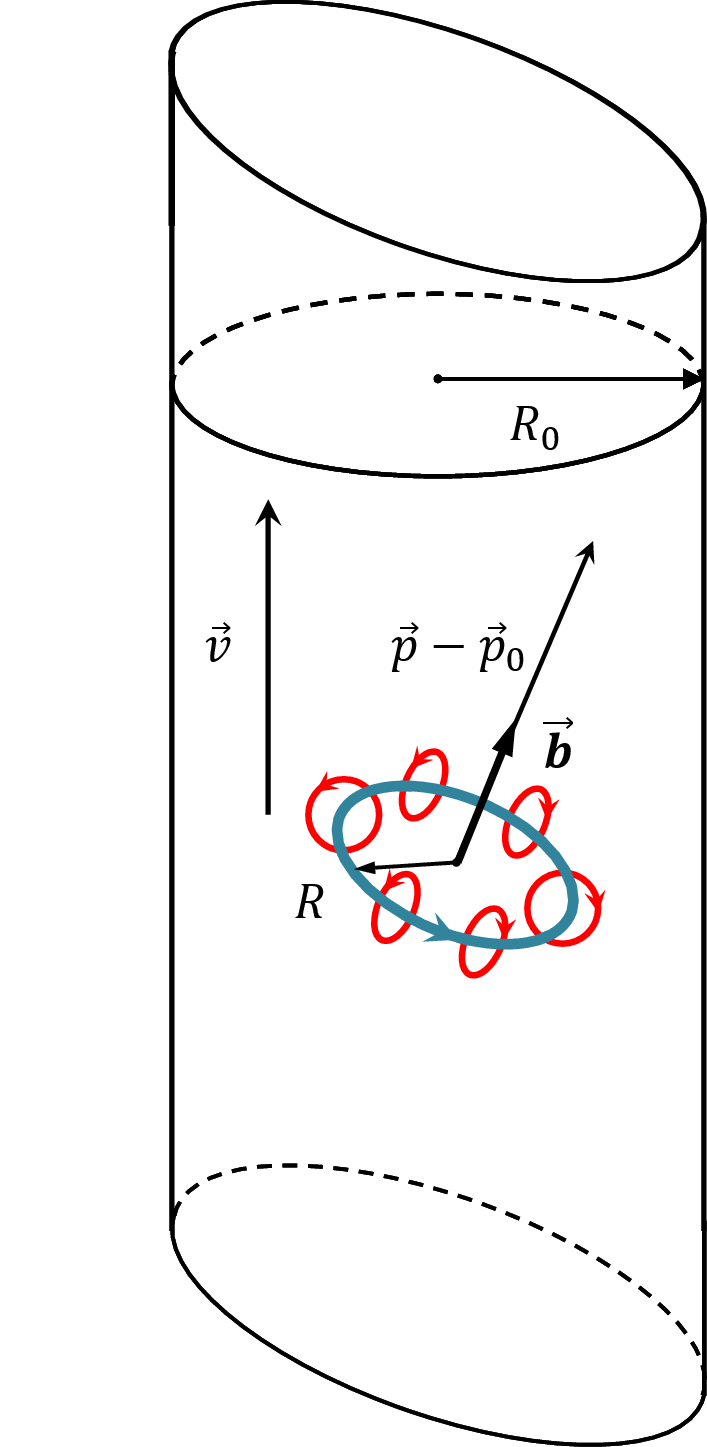}
{\vspace{15mm}\captionof{figure}{The circular vortex filament within the flowing fluid in a pipe.
\label{Pipe_1}}}
\end{center}

This approach means that we can avoid complicated and some ambiguous hydrodynamic calculations for vortex filaments with extremely thin cores, and we can determine the vortex energy in the following way\footnote{It is appropriate to recall Donnelly's remark here: {\it''\dots  considering how small the vortex core in helium II is, i.e., of order an angstrom, it would seem that one either ought to know how it is constructed or one ought to find a way to ignore it. Unfortunately neither goal has been achieved''} \cite{Donn}.	}:
		
\begin{equation}
		\label{E_general}
 {\cal E}_{cl} ~=~ \frac{{\bf p}^{\,2}}{2\mu_0} ~+~ {\hat C}_3(\varpi, \chi, \dots)\,.
\end{equation}
In this equation, the vector ${\bf p}$ stands for vortex momentum, while the variables $\varpi, \chi, \dots$ denote certain  ''internal'' variables.  
The explicit form of the function ${\hat C}_3$  depends on the type  of dynamics of those  variables.

Third, our choice of independent Hamiltonian variables allows us to use the formalism of many-body theory to describe vortex interactions. 
 {~  This area of study, vortex interaction, has long been intriguing (see, for example, \cite{Schwarz,KMD,KePoVe}).}
The author has previously addressed this topic, but it doesn't fit within the current paper's focus.

\subsection*{1. The main structures of the model}

This paper examines specific aspects of the circular vortex energy spectrum using the author's innovative approach.   First of all, we need to define a classical system that will be quantized.  
We consider an infinite cylindrical  pipe (see Fig.\ref{Pipe_1})   with a radius $R_0$, such that a fluid with a density $\varrho_0$ and speed of sound $v_0$ creates  laminar  flow  within this pipe. This  flow moves within pipe with a constant velocity $v$.

 In this pipe, a circular vortex filament with radius $R < R_0$ and small core radius $a$ moves such that the projection of its vortex momentum $\boldsymbol{p}$ on the axis of the cylinder is non-zero.


 Besides the constants $\varrho_0$ and $v_0$, which characterize the surrounding fluid, we also need to introduce a length constant $R_f$.  This constant can be related to the size of a stable molecular cluster, intermolecular distance and so on. Here, we don't specify it, but we assume that the constant $R_f > 0$ defines a certain minimal radius for the vortex ring.  
 Beside that, we will use the auxiliary constants such as 
$  t_0 = 2 R_0/v_0$  and  ${\mathcal E}_0 = \mu_0 v_0^2$
to simplify the formulas.  The  constants $t_0$ and  ${\mathcal E}_0$ define the  time and energy scales in constructed classical  theory. 
 
To describe the evolution of a circular vortex filament in the pipe, we use the local induction approximation  (LIA)  for thin filaments with  a non-zero   core  flow  \cite{AlKuOk}.
It is necessary to point out that the standard dynamical equation for this approximation writes for the evolution parameter $t^*=t\Gamma/4\pi$, where $t$  means  the actual (physical) time.
 
As geometrical object, this filament is described by the radius vector $\boldsymbol r(t, {\sf s})$, where $t$ is physical time and ${\sf s}$ is the natural parameter along the curve.
The suggested approach allows us to consider both small-perturbed circle-shaped vortex rings \cite{Tal22_1}   and loops of arbitrary size and shape \cite{Tal26_1}.
But in this study,  we only consider  circle-shaped rings  with a radius of $R$.  Exploring the dynamics of the vortex ring, it will be more convenient  to use dimensionless parameters $\tau$ and $\xi$ instead of the ''real time'' $t$ and the natural parameter ${\sf s}$:
\begin{equation}
        \label{tau_s}
 \tau    ~=~  t^* /  R^2\,,	\qquad \quad \xi ~=~ {\sf s}/{R}\,, \qquad
\xi \in [0,2\pi]\,.
 \end{equation}
If we introduce the projective vector  ${\mathfrak r} =  {\boldsymbol{r}}/R$, the convenient  form of the LIA equation will be as follows:

\begin{eqnarray}
        \label{LIE_pert}
        \partial_\tau {\mathfrak r}(\tau ,\xi)  & = &
        \alpha\, \Bigl(\partial_\xi{\mathfrak r}(\tau ,\xi)\times\partial_\xi^{\,2}{\mathfrak r}(\tau ,\xi)\Bigr)   ~+ \nonumber \\
				~~ & + & \omega\,\Bigl(2\,\partial_\xi^{\,3}{\mathfrak r}(\tau ,\xi) ~+~ 
        {3}\,\bigl\vert\, \partial_\xi^{\,2}{\mathfrak r}(\tau ,\xi)\bigr\vert^{\,2}\partial_\xi{\mathfrak r}(\tau ,\xi)\Bigr)\,,
				        \end{eqnarray}
where the values $\alpha$   and  $\omega$ are finite dimensionless constants.
In total, the parameter $\alpha$ arises as a consequence of a certain regularization process when the LIA equation is derived. As regards the
 parameter $\omega$, it  is determined by the velocity of the fluid flow  $\Phi_\omega$ in the vortex core.  A detailed deducing  of this equation { as well as deducing the explicit expression of the parameter $\omega$  from the conventional physical values} was made in  the book \cite{AlKuOk}.  Of course, the scope of application of LIA Equation  is quite limited.
To extend the applicability of Eq.(\ref{LIE_pert}), the author recently suggested its generalization \cite{Tal24_1}.

We will use the following  exact solution of Eq.(\ref{LIE_pert}): 

\begin{equation}
        \label{our_sol}
 {\mathfrak r}(\tau ,\xi) ~=~ \Bigl(\, {q_x}/{R} ~+~  \cos(\xi +\phi)\,,\quad {q_y}/{R} ~+~  \sin(\xi +\phi  )\,, \quad {q_z}/{R} ~+~ \alpha \tau \,\Bigr)\,. 
\end{equation}  
The variable  $\phi \equiv \phi(\tau) =  \phi_0 +  \omega\tau$. 
 determines  the rotation of the circular vortex filament (\ref {our_sol}) under the ring axis.
This fact means that variable  $\phi$  models the flow into the core of the vortex filament.
The variables $\boldsymbol{q} =
 (q_x, q_y, q_z)$ determine the position of the curve center in the  Cartesian coordinate system. 
In this study, we will be looking into this particular solution.

Beside the LIA equation (\ref{LIE_pert}), we also assume that 
  the conventional formula for the canonical momentum ${\boldsymbol p}$ is fulfilled: 
	  
   \begin{equation}
        \label{p_can}
        {\boldsymbol p} ~=~ {\boldsymbol p}_0 ~+~
				\frac{\varrho_0}{2 }\!\int\!\boldsymbol{r}\times\boldsymbol{w}(\boldsymbol{r})\,dV \,,
        \end{equation} 
	where the vorticity 	${\boldsymbol{w}}(\boldsymbol{r})$ is defined for the vortex filament by standard way:								
\[{\boldsymbol{w}}(\boldsymbol{r}) =  \Gamma\!\!
                  \int\limits_{0}^{2\pi}\!\delta(\boldsymbol{r} - \boldsymbol{r}(\xi))\partial_\xi{\boldsymbol{r}}(\xi)d\xi	\,.\]

	As regards to the vector $\boldsymbol{p}_0$ in  Eq.(\ref{p_can}), it reflects the presence of additional momentum that appears when the surrounding fluid moves at a non-zero velocity $\boldsymbol{v}$. 	
	 In our theory, we construct a non-relativistic system where the circular vortices are represented as particles with internal degrees of freedom. 
		In this context, we believe that we can use the standard relationship between  momentum $(\boldsymbol{P})$  and velocity $(\boldsymbol V)$  of a particle, which is $\boldsymbol{P} = {\sf M}\boldsymbol V$, where the symbol ${\sf M}$ represents some "mass". In our studies the fluid velocity $\boldsymbol v$ in the pipe is external parameter that is set "externally".

	Therefore,  we  postulate  that the following equality is fulfilled  for  additional vortex momentum ${\boldsymbol p}_0$:
	\begin{equation}
        \label{p_v_con}
				{\boldsymbol p}_0 ~=~ {\sf M}\boldsymbol{v}\,, \qquad {\sf M} ~=~ const\,,
			\end{equation} 	
		 Moreover, we  postulate that 	
		 the  constant  ${\sf M} = {\sf M_{eff}}$, where   the ''effective mass''   
		${\sf M_{eff}}$  of the vortex ring will be defined later (see Eq.(\ref{M_def})). 
	 We will return to the exact definitions and discussion of this issue in the following sections.
			{~ In fact,  equation (\ref{p_v_con}) establishes the connection between the additional vortex momentum ${\boldsymbol p}_0$, treated as an external parameter, and the actual external parameter $\boldsymbol{v}$.}
			The second summand  in Eq. (\ref{p_can})   is the standard  vortex canonical momentum ${\boldsymbol p}$ for zero boundary conditions $(v=0)$ at infinity \cite{Batche}.  

The function ${\boldsymbol r}(\tau ,\xi)$ describes a vortex filament as a geometrical object only. To take into account the motion of the surrounding fluid, we include the circulation $\Gamma$ as a dynamic variable in our theory. Therefore, the minimal set of variables needed to describe our dynamic system is as follows:
\[{\cal A} ~=~ \{ \Gamma, {\boldsymbol q}, R, \phi(\tau), {\boldsymbol b}\}\,,\]
 where the vector ${\boldsymbol b}$ means the binormal unit vector of the considered circular filament. Here, this vector coincides with the direction of the {~ symmetry axis of the vortex ring.}

If we substitute the solution (\ref{our_sol}) into Eq.(\ref{p_can}) we deduce   simple  equation
\begin{equation}
        \label{p_Gamma}
	{\boldsymbol{p}} ~-~ {\boldsymbol p}_0  ~=~    \pi\varrho_0 {R}^2 \Gamma\, {\boldsymbol b}\,. 
	\end{equation}

Our next step is the transformation of the set ${\cal A} $ into a new set ${ \cal A^{\prime}} $ which is more appropriate for quantization purposes.

 First, to achieve this goal, we make replacement 
$\{ \Gamma, {\boldsymbol b}\} \to  {\boldsymbol p}$
with help of the Eq.(\ref{p_Gamma}) where the vector ${\boldsymbol p}_0$ is considered as an external parameter.
Second,  we redefine the variables $ R$ and  $\phi(\tau)$ as follows:
\[ \chi  = \frac{\Delta R}{R_f}\cos(\phi_{\,0} +\omega\tau)\,, \quad  \varpi  = \frac{\Delta R}{R_f}\sin(\phi_{\,0} +\omega\tau) \,, 
\quad \text{where}\quad  \Delta R = \sqrt{R^2 - R_f^2}\,.\]
Clearly, the behavior of the variables $\varpi$ and  $\chi$  is similar to that of a harmonic oscillator.
Finally, we  postulate the set
\begin{equation}
\label{new_set1}
{\cal A}^{\,\prime} ~=~ \bigl\{\, {\boldsymbol p}\,,  {\boldsymbol{q}}\,;\ \varpi\,, \chi\,
  \,\bigr\}
 \end{equation}
as the set of fundamental independent variables for the considered dynamical system -- circle-shaped vortex ring which evolves in accordance with Eq.(\ref{LIE_pert}).
As a consequence of equality (\ref{p_Gamma}), the following formula holds:
\begin{equation}
\label{main_con}
({\boldsymbol p} ~-~ {\boldsymbol p}_0)^{\,2}   ~=~  \pi^2\varrho_0^2 {R_f}^{\,4}\Bigl(1 + \varpi^{\,2} + \chi^{\,2} \Bigr)^{2}\, \Gamma^{\,2}\,.
\end{equation}
This formula shows that the circulation $\Gamma$ is the function of new variables that parametrize the set (\ref{new_set1}). 

In general, the definition of the Hamiltonian structure of a non-linear dynamical system such as system (\ref{LIE_pert})  is a complex problem. The main steps of the corresponding theory can be found in the book by \cite{TakFad} (see also work \cite{Sasaki}, where this issue was investigated from the perspective of differential geometry).
  In our specific scenario, the Hamiltonian structure emerges naturally from the structure  of set ${\cal A}^{\,\prime}$:

	\begin{itemize}
  \item Our system's phase space is the product of two spaces: ${\mathcal H} =  {\mathcal H}_{pq}  \times  {\mathcal H}_b   $. The space $\mathcal{H}_{pq}$ serves as the phase space for a three-dimensional, free, structureless particle. This space is defined by the variables ${\boldsymbol{q}}$ and ${\boldsymbol{p}}$.  The space    $ {\mathcal H}_b$  is a phase space for one-dimensional harmonic oscillator with  frequency $\omega/t_0$.
		 \item Poisson structure (brackets that are not written are vanished):
  \begin{equation}
  \{p_i\,,q_j\}  =  \delta_{ij}\, \quad (i,j = x,y,z)\,, 
  \qquad 
  \{  \varpi,\, \chi\}  =  {1}/{\cal E}_0 t_0 \nonumber 
  \end{equation}
  	\item Hamiltonian 
\begin{equation}
        \label{hamilt_1}
				H^\# ~=~ \frac{\boldsymbol{p}^2}{2 \mu_0} ~+~  \frac{{\cal E}_0\,\omega}{2}\Bigl(\varpi^2  + \chi^2  \Bigr)\,.
	\end{equation}
	\end{itemize}
This form of the function $H^\#=H^\#({\boldsymbol p}\,;\ \varpi\,, \chi\,)$ is a direct consequence of our definition of energy -- see Eq.(\ref{E_general}).  

We must emphasize here that the Hamiltonian $H^\#$ provides dynamics for variables in accordance with conditional time $t^\# = t_0 \tau$. For example,

\begin{equation}
e^{-t^\#\{H^\#,\dots }\,\chi(0) =  \chi(t^\#) ~=~ \frac{\Delta R}{R_f}\cos\left(\phi_{\,0} +\frac{\omega\, t^\# }{t_0}\right) \,,\nonumber \\[2mm]
\end{equation}
and so on. In the final formulas, we need to fulfill  the procedure of returning  $t^\# \to t$    to physical time $t$. 

{~
Let's note the pertinent detail. 
We declared the physical quantity $\Gamma$ to be an additional dynamic variable.
This extension, as well as subsequent Hamiltonian structure construction  of our model, makes the dynamics of the vortex ring more diverse compared to those described by Eq.(\ref{our_sol}). Indeed, our Hamiltonian structure leads to dynamics
\[{\boldsymbol q}(0) ~\longrightarrow~ {\boldsymbol q}(t^\#) ~=~ {\boldsymbol q}(0) ~+~ \frac{\boldsymbol p}{\mu_0}\,t^\#\,,\]
but the Eq.(\ref{our_sol}) provides the dynamics
\[ {\boldsymbol q}(0) ~\longrightarrow~ {\boldsymbol q}(t^\#) ~=~ {\boldsymbol q}(0) ~+~ 
		\alpha{R}\,(t^\#/t_0)\,{\boldsymbol b}\, .\]  }

\subsection*{2. Quantization}

The above construction naturally leads to a quantum description of our vortex filament in terms of  the Hilbert space
\begin{equation}
	\label{space_quant}
	\boldsymbol{H}_1  ~=~  \boldsymbol{H}_{pq} \otimes   \boldsymbol{H}_b  \,.
	\end{equation}
			The symbol   $\boldsymbol{H}_{pq}$  denotes here the Hilbert space  of a free structureless particle in the space domain  $\mathbb{D}_3$, where the symbol  $\mathbb{D}_3$ means the infinite circular pipe described above.  In this study we assume 			$\boldsymbol{H}_{pq} = L_2(\mathbb{D}_3)$.  	The symbol $\boldsymbol{H}_b$ represents the Hilbert space of the quantized harmonic oscillator, characterized by classical variables $\chi$ and $\varpi$. 
				This space is formed by the vectors
		\[|\,n\rangle   ~=~  \frac{1}{\sqrt{n!}} (\hat{b}^+)^n    |\,0_b\rangle   \qquad 
		[\,\hat{b}, \hat{b}^+] ~=~ \hat{I}_b\,, \quad \hat{b}|\,0_b\rangle ~=~ 0\,,  \]
			where vector  $|\,0_b\rangle \in \boldsymbol{H}_b$ is vacuum vector and symbols $\hat{b}^+$ and $\hat{b}$ mean the  operators which create and annihilate the oscillator levels. 	
We quantize the  variables  $\varpi$ and $\chi$  in the following way:

\begin{equation}
	\label{quant_rules}
 \chi ~\to~ \sigma_{ph}\, \frac{\hat{b} + \hat{b}^+}{\sqrt{2}}\,, \qquad 
\varpi ~\to~ \sigma_{ph}\, \frac{\hat{b} - \hat{b}^+}{i\sqrt{2}}\,, \qquad \sigma_{ph} ~=~ \sqrt{\frac{\hbar}{{\mathcal E}_0 t_0}}\,.
\end{equation}
Here, we introduced  the dimensionless constant $\sigma_{ph}$ which 
depends on the Planck constant $\hbar$. 
The variables $\mathbf{q}$ and $\mathbf{p}$ are quantized using the standard rules of quantum mechanics using a coordinate representation. This applies to a free, non-relativistic particle within the Hilbert space $L_2(\mathbb{D}_3)$.

The quantization scheme leads to  two spectral problems.
\begin{itemize}
\item  The first  spectral problem appears after quantization of the equality (\ref{main_con}). This problem defines the circulation values $\Gamma$:
\begin{equation}
\label{eq_sp_Gamma}
  \hspace{-5mm}
\left[ \bigl(-i\hbar\partial/\partial z - p_0\bigr)^2  - \hbar^2\Delta_2  - \pi^2 \varrho_0^2 \Gamma^2 R_f^4  \Bigl({\hat I}  +  \sigma_{ph}^2\,
(\hat{b}^+ \hat{b} + 1/2)  \Bigr)^2\right]|\Psi\rangle = 0\,.
	\end{equation}
	Here and further,  the symbol $\Delta_n$ stands for the Laplace operator in the  $L_2(\mathbb{R}_n)$ space.
\item Second problem is determining the energy value $E^\#$, such that the corresponding Hamiltonian
   $\widehat{H}^\#$ provides dynamics in accordance with conditional time $t^\# =  \tau t_0$:
\begin{equation}
	\label{H_quant}
	\widehat{H}^\# |\psi\rangle = E^\#|\psi\rangle\,, \qquad
		\widehat{H}^\#  ~=~ -~  \frac{\hbar^2}{2\mu_0}\Delta_3 ~+~ \frac{\hbar\, \omega}{t_0}
		\left(b^+ b + \frac{1}{2}\right)\,.
		\end{equation}
\end{itemize}
The spectral problems  (\ref{eq_sp_Gamma})  and  (\ref{H_quant})  have a common set of the eigenvectors:

\begin{equation}
		\label{R_state}
	 \boldsymbol{H}_1^\prime ~\ni~     |\psi_{{p};\ell, m;n}\rangle ~=~ 
	|\,{p}; \ell, m\rangle|n\rangle\,, \qquad\quad
		|n\rangle  \in \boldsymbol{H}_b\,,
		\end{equation}
		where  the vector $|\,{p}; \ell, m\rangle \in  \boldsymbol{H}_{pq}^\prime$  is the eigenvector of the  operator $\Delta_3$.  
				Symbol $\boldsymbol{H}_{pq}^\prime$ means corresponding rigged (''equipped'') Hilbert space (see, for example, \cite{BerShu}). We also assume that the corresponding wave functions vanish on the surface of the pipe in coordinate representation.
				In this  representation, vectors  $|\,{p}; \ell, m\rangle$ take form
		\[ \langle\,\boldsymbol{q}\,|\,{p}; \ell, m\rangle ~=~ \exp({i pz}/{\hbar})\exp(i\ell\varphi)
		J_\ell(\zeta^{(\ell)}_m \rho/R_0)\,,\]
where the values $(\rho, \varphi, z)$ represent the cylindrical coordinates of the vector $\boldsymbol{q}$,  where the $Z$-axis is directed along the pipe axis.
The numbers
$\zeta^{(\ell)}_m$ mean $m$th zero for the Bessel function $J_\ell$. To simplify subsequent formulas, we further assume that the number $\ell=0$.

Corresponding  eigenvalues
	$\Gamma = \Gamma_{m}(p)$ for the spectral problem (\ref{eq_sp_Gamma})  will be like this:
		
\begin{equation}
\label{Gamma_val}
\Gamma_{m,n}(p) ~=~  \pm \frac{\hbar}{{\sf M}_0} 
 \frac{\sqrt{(\kappa - \kappa_0)^2  +(\zeta^{(0)}_m)^2 }}{ \bigl[ 1 +  
\sigma_{ph}^2(n + 1/2)\bigr]}\,, \qquad  
 m = 1,2,\dots.,
\end{equation}
where the ''mass'' constant ${\sf M}_0 = \pi \varrho_0  R_f^2R_0$ 
 was introduced. The variables $\kappa = p/p_{\hbar}$ and 
$\kappa_0 = p_0/p_{\hbar}$ are dimensionless in accordance with  definition of the momentum scale constant $p_{\hbar}   = \hbar/R_0$.

The eigenvalues $E^\#_{m,n}(p)$  for the spectral problem (\ref{H_quant}) are like this:
\begin{equation}
	\label{E_eigen1}
	E^{\#}_{m,n}(p) ~=~ \frac{p_{\hbar}^2 }{2\mu_0 }  \Bigl(\kappa^2 + (\zeta^{(0)}_m)^2  \Bigr) ~+~
	\frac{\hbar\,\omega}{t_0}\, \left(n + \frac{1}{2}\right)\, \qquad n = 0,1,\dots\,.
	\end{equation}

	Next, we must remember that until now we have been describing the LIA-dynamics in terms of special evolution parameter $t^*=t\Gamma/4\pi$.   				
		After that, we introduced   the dimensionless evolution parameter $\tau$  in accordance with Eq.(\ref{tau_s}) for convenience. 
		Taking into consideration the  constructed Hamiltonian structure of our model, 		
		this fact means  as a pass to  evolution in terms of  the  ''conditional time'' $t^\# = t_0\tau$.  
		Therefore, the energy that determined by the Eq.(\ref{E_eigen1}) plays a role as some ''conditional energy''.		This fact should be taken into account because real energy is a physical quantity that is associated with real time translations.

				
				Thus, the $t^\#$-dependence 
 of any vector $|\Psi\rangle \in {\boldsymbol H}$ is described in the standard way:
		\begin{equation}
	\label{t_ev1}
		\exp\left(\frac{i\widehat{H}^\# t^\#}{\hbar} \right)|\Psi\rangle ~=~  
		\sum_{m,n} \int\!\! dp\, C_{m,n}(p)\exp\left(\frac{iE^\#_{m,n}(p) t^\#}{\hbar} \right)
		|\psi_{{p};0, m;n}\rangle\,,
		\end{equation}
	where  $ C_{m,n}(p) = \langle\psi_{{p};0, m;n}|\Psi\rangle$.	In the next step,
					we  restore the real-time $t$ value in our formulas by using the equation
				\begin{equation}
	\label{t_real}
		 t^\# ~\to~  t ~=~ \frac{4\pi R^{\,2}}{ t_0 |\Gamma|}\,t^\#\,.
		\end{equation}
							
		{~ In the quantum case, where $R$ and $\Gamma$ are quantized, a hierarchy of temporal scales emerges. When we will consider a many-vortex system, the temporal scale  can change between one vortex and another. A spatio-temporal hierarchy within turbulent motion was discovered long ago in classical hydrodynamics \cite{Kolm}.  Thus, suggested approach gives new insights on this important  phenomenon on a quantum level.  
	The author devoted a separate paper to this issue \cite{Tal_PhScr24}, which is why we will not discuss the details in this paper.	}

				To restore the actual energy, 	let us note that the amplitudes
						$\langle \Phi  |\Psi(t^\#)\rangle $ of transitions  to any quantum  state $|\Phi\rangle$ determine  an observable values. Thus, they should not depend on time parametrizations.				To ensure this, we need to  require 	the following equality  for the real energy $E_{m,n}(p)$: 									
				\[  E^\#_{m,n}(p) t^\#  ~=~  E_{m,n}(p) t\,.\]
				Therefore, 				the ''real - time'' evolution of any vector $|\Psi\rangle \in {\boldsymbol H}$   is written as
		\begin{equation}
	\label{t_ev2}
		|\Psi\rangle ~\longrightarrow~ |\Psi(t)\rangle ~=~  
		\sum_{m,n} \int\!\! dp\, C_{m,n}(p)\exp\left(\frac{iE_{m,n}(p)\, t}{\hbar} \right)
		|\psi_{{p};0, m;n}\rangle\,,
		\end{equation}
					where function $E_{m,n}(p)$ means the actual vortex energy. 
					{~ Thus, our energy calculation process can be outlined as follows:
					
		\vspace{3mm}

		{\Large
\[
\begin{array}{ccccc}
\framebox(55,20){ \scalebox{0.8}{{\sf LIA}}} &\hspace{-3mm}\xLongrightarrow[\text{~~~}]{\text{${\sf{t}\to \sf{t}^\#}$}} & \framebox(95,20){\scalebox{0.7}{\sf Dynamic Eq.(\ref{LIE_pert}) }} & {\not\Longrightarrow}  & ~~E_{m,n}(p)  \\[3mm]
 \Bigg\Uparrow 
\lefteqn{\begin{turn}{90}${\hspace{-8mm}{~\sf{a} \simeq 0}}$\end{turn}}
&& \Bigg\Downarrow\lefteqn{\begin{turn}{-90}{\hspace{-12mm}\scalebox{0.7}{ \sf hamilt. str.}}\end{turn} }\vspace{3mm}&&\Bigg\Uparrow
  \lefteqn{\begin{turn}{90}${\hspace{-8mm}{\sf{t}^\# \to \sf{t}}}$\end{turn}}\\[5mm]
	\framebox(75,20){ \scalebox{0.8}{{\sf Vortex ring}}} && \framebox(95,20){ \scalebox{0.7}{ $H^\#:$ ~\sf Eq.(\ref{hamilt_1})}} & \xLongrightarrow[\text{\sf quantization}]{\text{${\hbar}$}}\hspace{-4mm}  & ~~E^{\#}_{m,n}(p) 
    \end{array}
\]} }
		
		\vspace{1mm}	
		
	{\em Diagram:}   Illustrates the sequence of theoretical steps leading to the derived energy dependence $E_{m,n}(p)$.

	\vspace{5mm}

			Finally, the function  $E_{m,n}(p)$  takes the following form:
\begin{equation}
	\label{E_true}
	E_{m,n}(p)   ~=~ {~ \frac{t_0 |\,\Gamma_{m,n}(p)\,|}{4\pi R^2_n}  E^\#_{m,n}(p)  } ~\equiv~ E_{m,n}(p; p_0)	~=~  E_{\hbar}\,{\cal E}_{m,n}(\kappa, \kappa_0)\,,
	  \end{equation}
	where the constant $ E_{\hbar} ~=~ {\hbar^2}/{4\pi {\sf M}_0 R_f^2}$
	defines the energy scale  in our model. The dimensionless function ${\cal E}_{m,n}(\kappa; \kappa_0)$
	is written as follows:
	\begin{equation}
	\label{E_dim_less}
{\cal E}_{m,n}(\kappa, \kappa_0) ~=~ \frac{\sqrt{(\kappa - \kappa_0)^2  +(\zeta^{(0)}_m)^2 }}{ \bigl[ 1 +  
\sigma_{ph}^2(n + 1/2)\bigr]^2}\left[ \omega\bigl(n + 1/2\bigr) + \sigma_{ph}^2 \bigl(\kappa^2 + (\zeta^{(0)}_m)^2 \bigr)\right]\,.
\end{equation}

We want to note here that the energy is finite even in the case of $p = 0$ and ${p_0} = 0$.  In the paper \cite {MarPar}, this property of the vortex spectrum was stated as a necessary condition  when the vortex branch is considered as a continuation of the  phonon-roton  branch in Helium II.

\subsection*{3. Energy stationary points and the $p_0 (v)$ dependency}

For our upcoming research, it's crucial to compare the dependency $E\nolinebreak = \nolinebreak E_{m,n}(p;p_0)$ for various $p_0$ values, keeping $m$ and $n$ constant.  For simplicity's sake, let's assume $p_0 \ge 0.$  
The stationary points of the function ${\cal E}_{m,n}(\kappa, \kappa_0)$ are defined by
 the equation
\[ \frac{\partial{\cal E}_{m,n}(\kappa, \kappa_0)}{\partial\kappa} ~=~ 0\,,\]
which is reduced to the cubic {~ algebraic} equation. 
{~ When the  discriminant $\Delta$ of this equation takes negative value, corresponding cubic equation has a three real-valued roots and has one real-valued root when $\Delta > 0$.}
Thus, the number of real stationary points  $p_i(p_0;m,n)$ of the function 
${\cal E}_{m,n}(p; p_0)$ can changes from one  ($p_{main}$)  to three ($p_1, p_2,  p_{main}$), depending on the value of $p_0$.
The visible result is represented in Fig.\ref{Ep_dep} {~  for $\omega = 10^{-9}$
 and $\sigma^2_{ph} = 10^{-6}$.}


\begin{center}
		 {\includegraphics[width=5.5in]{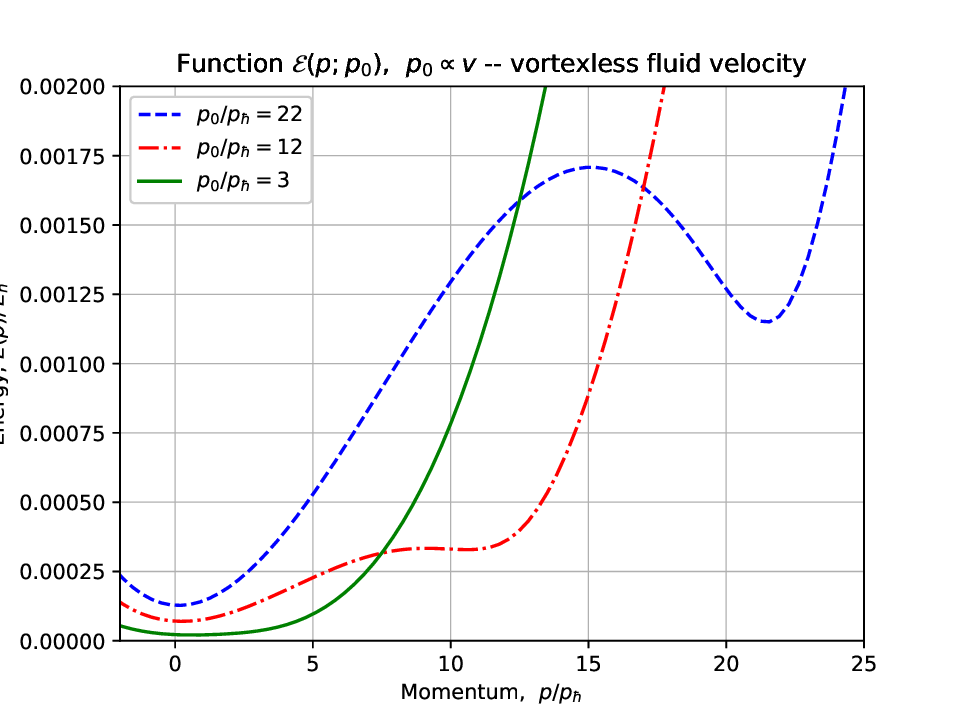}}
{\captionof{figure}{ The transformation of the function $ E = E_{1,1}(p; p_0)$ 
 with respect to the parameter  $p_0$ ~~($\omega = 10^{-9}$).
\label{Ep_dep}}}
\end{center}


These graphs demonstrate a non-trivial effect: the increase of the additional momentum $p_0$ leads to the appearance of additional stationary points in the function being considered. 
These points can correspond to both local minima and local maxima of the function ${\cal E}_{m,n}(p; p_0)$ as well as the flex point.

Let us study the connection (\ref{p_v_con})  between the  additional momentum $p_0$ and the vortexless fluid velocity $v$ in more detail.
At the outset, we propose a definition for vortex effective mass
\begin{equation}
	\label{M_def}
	({\sf M_{eff}})^{-1} ~=~  \frac{\partial^{\,2} E_{m,n}(p;p_0)}{\partial p^{\,2}}\,\bigg\vert_{\,p ~=~ p_{\,\sf{extr}}(p_0;m,n)}\,,
\end{equation}
 where the value $p_{\sf extr} = p_{\sf extr}(p_0;m,n)$ means one of the stationary points of the function $E_{m,n}(p;p_0)$.

\begin{center}
		 {~\includegraphics[width=5.5in]{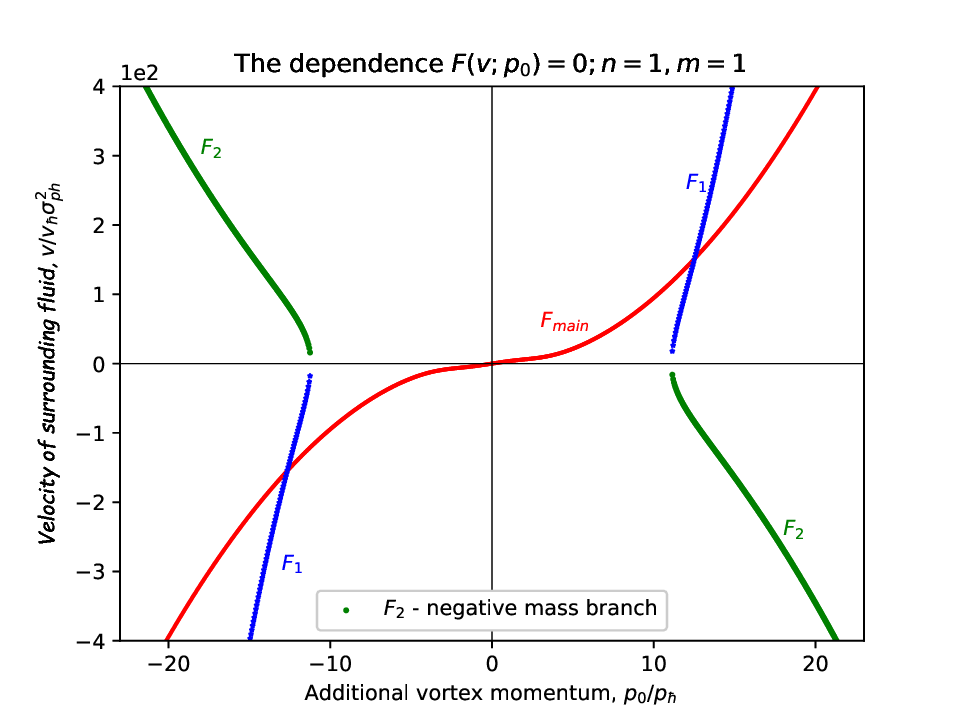}}
{\captionof{figure}{Dependency  $v = p_0\,{\sf M^{-1}_{eff}}(p_0)$
 for the different selection  of the effective mass, ~~$\sigma^2_{ph} = 10^{\,-6}$,   ~~$\omega = 10^{\,-9}$ .
\label{p_v_dep10}}}
\end{center}

Thus, the Eq.(\ref{p_v_con})   generally  takes the following     form:
 \[v ~=~  p_0 \frac{\partial^{\,2} E_{m,n}(p;p_0)}{\partial p^{\,2}}\,\bigg\vert_{\,p ~=~ p_{\,\sf extr}(p_0;m,n)}\,.\]

 The different choice of stationary point $p_{\sf extr}$ in this equation gives different dependencies $F(v,p_0) = 0$.  Earlier in this section, we mentioned that the number of these points can vary from one to three. 
Al possible dependencies are shown on Fig.\ref{p_v_dep10} graphically.

 $\star$  The following definitions of the graphs $F_{main}$ (red curve), $F_{1}$ (blue curve) and 
$F_{2}$ (green curve) take place:

\begin{equation}
	\label{p_v_con2}
	 F_{main}: \quad \text{graph of the function} \quad
v ~=~  p_0 \frac{\partial^{\,2} E_{m,n}(p;p_0)}{\partial p^{\,2}}\,\bigg\vert_{\,p ~=~ p_{main}(p_0;m,n)}\,,
\end{equation}

\begin{equation}
	\label{p_v_con3}
		F_{i}:\quad \text{graphs of the functions} \quad
v ~=~  p_0 \frac{\partial^{\,2} E_{m,n}(p;p_0)}{\partial p^{\,2}}\,\bigg\vert_{\,p ~=~ p_{i}(p_0;m,n)}\,,
\end{equation}
where $i = 1,2$.
In these graphs, we conveniently use the coordinates: the ''dimensionless momentum'' $p/p_{\hbar}$ and the ''dimensionless velocity'' $v/v_{\hbar}$, where $v_{\hbar} = E_{\hbar}/p_{\hbar}$.

So, we have a single stationary point for the sufficiently small values  of the  momentum $p_0$. Further, when this parameter increases, the flex point as well as additional stationary points appear on the graph ${\cal E} = {\cal E}_{m,n}(p; p_0)$ (see Fig.\ref{Ep_dep}). 

The appearance of additional stationary points, both local minimum in the point $p_1 (p_0; m, n)$ and local maximum in the point $p_2 (p_0; m , n )$ correspond to additional quasi-stable states of the vortex, which have positive effective mass at the point $ p_1 ( p_0; m, n)$ and negative effective mass at the point $p_2 ( p_0; m, n)$.

We emphasize here that the function $F(v;p_0) = 0$ is shown for case $v \not= 0$ only.
The trivial case of $v=0$ must be considered separately, when the value of $p_0=0$  in the Eq.(\ref{p_can}). Therefore, the points of $(p_0\not=0, v =0)$ on the graph in  Fig.\ref{p_v_dep10}
correspond to the limit $v \to 0$, $v \not= 0$. This limit means the appearance of a flex point on the graph in Fig.\ref{Ep_dep}.
Formally speaking, this flex-point corresponds to the infinitely large effective mass. 
The case $v \simeq 0$,  $v \not= 0$ corresponds to a vortex with an extremely large effective mass.
Probably, similar vortex states could be useful in cosmological models\footnote{See, for example, the paper \cite{Nature}, where a certain vortex model is applied to describe curved space-time.}.
In systems with macroscopic scales but not cosmological dimensions, the presence of such single vortices appears questionable.

\subsection*{4. Discussions}

~~{~ Formally speaking, our constructions lead to vortices of infinite mass.  Indeed, in Fig.\ref{p_v_dep10}, the junction of graphs $F_1$ and $F_2$ marks the point where the second derivative of $E$ with respect to $p$ in Eq. (\ref{p_v_con3}) becomes zero.   Of course,  these states are unphysical. Next, we'll explore the restrictions that help prevent the appearance of such states.  To make it, we need to consider vortex pairs, and not just single vortices. It is possible to do this, because }  our approach can  easily be   expanded on the many vortex system\footnote{~Despite this item lying outside of our study, we will make some remarks about it in conclusion.}. So, we can consider the Hilbert space $\boldsymbol{H}_2 =  \boldsymbol{H}_1 \otimes\boldsymbol{H}_1$ to describe the pair of non-interacting  or weakly interacting vortices. In this case, each  vortex $''i''$  will  move in a fluid with velocity flow of $\boldsymbol{v} +  \boldsymbol{\delta v}_i$, where $\boldsymbol{\delta v}_i$ is the perturbation that caused by another vortex.  In the case where $|\,\boldsymbol{\delta v}_i\,| << v$, the energy specrta for both vortices are the same. Indeed,  the inequality of $|\,{\delta p}_i\,| \not= 0 $ is always true due to the Heisenberg Uncertainty Principle.
Although this issue needs further research, we can make some estimates based on our models.

Yet, the summ mass of a vortex pair, comprising a vortex with positive effective mass ${\sf M^{(+)}_{eff}}$ and another with negative effective mass ${\sf M^{(-)}_{eff}}$, remains finite even when  the velocity $v$ approaches zero!
Indeed, let us perform  the Taylor expansion of the function ${\cal E}(\kappa)$ (see Eq.(\ref{E_dim_less})  and Fig.\ref{Ep_dep}) around the critical  point $\kappa_c = p_c/p_{\hbar}$:
{~ \[ {\cal E}(\kappa) ~=~ {\cal E}(\kappa_c)  ~+~
\frac{1}{3!}{\cal E}^{(III)}(\kappa_c)(\kappa - \kappa_c)^3  ~+~
\frac{1}{4!}{\cal E}^{(IV)}(\kappa_c)(\kappa - \kappa_c)^4  ~+~ \dots \]  }
 In this point, the branches $F_1$ and $F_2$ connect to each other  (see Fig.\ref{p_v_dep10}). 
As the consequence, we conclude:

\begin{eqnarray}
\label{est_1}
 \lim_{\varepsilon \to 0}\,\Bigl[ {\sf M^{(+)}_{eff}}(p_c - \varepsilon)   ~+~ 
   {\sf M^{(-)}_{eff}}(p_c   + \varepsilon)\Bigr]  &=&\\[2mm]
~=~ \frac{(p_{\hbar})^2}{E_{\hbar}}  \lim_{\varepsilon \to 0}
\,\biggl[\biggl(\frac{d^2 \cal E}{d\kappa^2}\biggr)^{-1}\bigg\vert_{\kappa = \kappa_c -\varepsilon} ~+~  \biggl(\frac{d^2 \cal E}{d\kappa^2}\biggr)^{-1}\bigg\vert_{\kappa = \kappa_c +\varepsilon} \biggr] &=&
-~ \frac{(p_{\hbar})^2}{E_{\hbar}} \frac{{\cal E}^{(IV)}(\kappa_c)}{[{\cal E}^{(III)}(\kappa_c)]^2}\,.  \nonumber
\end{eqnarray}
 The symbols ${\cal E}^{(\dots)}(\kappa_c)$
denote the corresponding  derivatives which are  calculated in the 
 critical point $\kappa_c$.

 What do these facts mean? First of all, it should be noted that fluid velocity $v$ is the only control parameter that can be directly changed by external conditions.
In accordance with the graphs on in the Fig.\ref{p_v_dep10}, the meta-stable states can arise for 
every value of the velocity $v$.  Let us assume that this parameter takes the value
\[ v ~\in~ (0, \infty)\,.\]
The above analysis shows that small velocities $v$ correspond to vortices with small momenta
 $p_0$  (the branch $F_{main}$ on Fig.\ref{p_v_dep10}), as well as extra vortex pairs consisting of vortices of positive and negative mass with momenta $p_0 > p_c$.
Therefore, the evolution of a circular quantum vortex filament in a flowing fluid can give rise to a coupled vortex pair. This pair includes a vortex with positive effective mass, $\sf M^{(+)}_{eff}$, and a vortex with negative effective mass,  $\sf M^{(-)}_{eff}$.
{~ Let us note that states which exhibit the properties of particles with negative mass have been studied by various authors (see, for example, \cite{CoLaDa,SakMal}).}
In our opinion, this event can be interpreted in a flowing fluid as the physical effect that contributes  to the transition to  the turbulent regime.

Let us study the additional restrictions imposed by quantum theory on the value of $v$ in this context. 
To  demonstrate this, we note that the pure coordinate states $|\boldsymbol{q}\rangle \in
\boldsymbol{H}_{pq}^\prime$ as well as the states $|\boldsymbol{p}\rangle \in
\boldsymbol{H}_{pq}^\prime$  are convenient for the formulation of the theory but unphysical.
Instead the states $|\boldsymbol{q}\rangle$ and $|\boldsymbol{p}\rangle$  we will use the coherent states $|\boldsymbol{z}\rangle$    for the non-relativistic massive particle \cite{BagGit}. Such states for the description of vortex filaments of arbitrary shape and size are applied by the author in the paper \cite{Tal26_1}.  It is known that these states minimize the Heisenberg uncertainty relations.
We also suppose that the relations
\[\delta x  ~\approx~ \delta y ~\approx~ \delta z\,, \qquad
\delta x = \sqrt{\overline{(x - \overline{x})^2}}\,, \quad \delta y = \dots\,  \]
hold, where  symbol $\overline{x}$ means averaging over coherent states  $|\boldsymbol{z}\rangle$.

Furthermore, it is natural to assume that the uncertainty $\delta z$ at a spatial point where a vortex forms is of the order of $R_0$.
In this context, the total momentum $p_s$ of the created coupled vortex pair is approximately equal to the uncertainty  $\delta p \simeq p_{\hbar}$.  Indeed, momentum $p_s$ can be estimated using the following formula:
\begin{equation}
	\label{ps_est}
 p_{s}  ~\equiv~ \Big\vert\,p^{(+)}  ~+~  p^{(-)}\,\Big\vert ~\simeq~ \delta p 
 ~\simeq~   \hbar/R_0  ~=~  p_{\hbar}\,,
\end{equation}
where
\[ p^{(+)} ~=~ p^{(+)}_{\Gamma} ~+~ {\sf M^{(+)}_{eff}}(n,m)v\,, \qquad 
p^{(-)} ~=~ p^{(-)}_{\Gamma} ~+~ {\sf M^{(-)}_{eff}}(n,m)v\,.\]

The terms $p^{(+)}_{\Gamma}$ and  $p^{(-)}_{\Gamma}$ correspond to the second summand in the general momentum equation (\ref{p_can}). These elements, along with circulation $\Gamma$, can have any signs and values.  Because the inequality
\[\Big\vert\,\Bigl(p^{(+)}_{\Gamma} ~+~ {\sf M^{(+)}_{eff}}v \Bigr)  ~+~  \Bigl(p^{(-)}_{\Gamma} ~+~ {\sf M^{(-)}_{eff}}v \Bigr)\,\Big\vert ~\le~
\Big\vert\,p^{(+)}_{\Gamma} ~+~ p^{(-)}_{\Gamma}\,\Big\vert   ~+~ \Big\vert\,{\sf M^{(+)}_{eff}}
v   ~+~   {\sf M^{(-)}_{eff}}v \,\Big\vert\,,\] the estimation (\ref{ps_est}) is transformed as follows:
\begin{equation}
	\label{v_est0}
	\Big\vert\,{\sf M^{(+)}_{eff}}   ~+~   {\sf M^{(-)}_{eff}} \,\Big\vert v ~\ge~ \hbar/R_0 ~-~
	\Big\vert\,p^{(+)}_{\Gamma} ~+~ p^{(-)}_{\Gamma}\,\Big\vert \,.\nonumber
\end{equation}
This inequality will be true for all values $p^{(+)}_{\Gamma}$ and  $p^{(-)}_{\Gamma}$ if the velocity $v \ge v_{min}$, where the value $v_{min}$ satisfies equality
\begin{equation}
	\label{v_est}
	v_{min}\max\Big\vert\,{\sf M^{(+)}_{eff}}   ~+~   {\sf M^{(-)}_{eff}} \,\Big\vert  ~\simeq~ \hbar/R_0  \,.
\end{equation}
 
The maximum of the value $\Big\vert\,{\sf M^{(+)}_{eff}}   +   {\sf M^{(-)}_{eff}} \,\Big\vert$
is achived in the critical point $p_c$. The Figure \ref{M_limits} visually demonstrates this fact.

Finally, we can state a criterion that must be fulfilled  for the initiation of the processes of creation of the coupled vortex pair in question:

\begin{equation}
\label{criterion}
{\sf Re}_q ~\equiv~
\frac{\varrho_0 v R_f^4}{\hbar} ~\ge~ \frac{[{\cal E}^{(III)}(\kappa_c)]^2}{4\pi {\cal E}^{(IV)}(\kappa_c)} ~\equiv~ {\mathfrak R}_q^c\,.
\end{equation}

The value ${\sf Re}_q$ is interpreted as the quantum Reynolds number for our model.  Besides the flow velocity $v$, it only depends on the constants of the fluid. 
The constant ${\mathfrak R}_q^c$  in the right hand side of the inequality (\ref{criterion}) depends on the kind of the domain where the fluid flows. Moreover, this constant depends on the constants $\omega$    and  $\sigma_{ph}$ (see Eq.(\ref{quant_rules}))  as well as the quantum numbers $n$ and $m$. These constants and numbers are attribute  to the initial circular vortex ring that corresponds to the Branch $F_{main}$ in Fig. \ref{p_v_dep10}.
We would like to note that criterion (\ref {criterion}) is only valid in a neighborhood of velocity $v_ {min}$. The case of $v \to \infty$ requires of additional studies.

 We should note that all the graphs shown above crucially depend on the parameters $\omega$ and $\sigma_{ph}$. The values of critical points and intersection points may change dramatically, as well as other significant features.   
 Because the actual values of these parameters require subsequent corrections (both theoretical and experimental), we were guided primarily by considerations of visibility here. For example, Fig.\ref{Ep_dep2} shows the changes in the
 Fig.\ref{Ep_dep},  when the  constant $\omega$ is changed from  $\omega = 10^{\,-9}$ to $\omega = 10^{\,-3}$.

\begin{center}
		 {\includegraphics[width=5.0in]{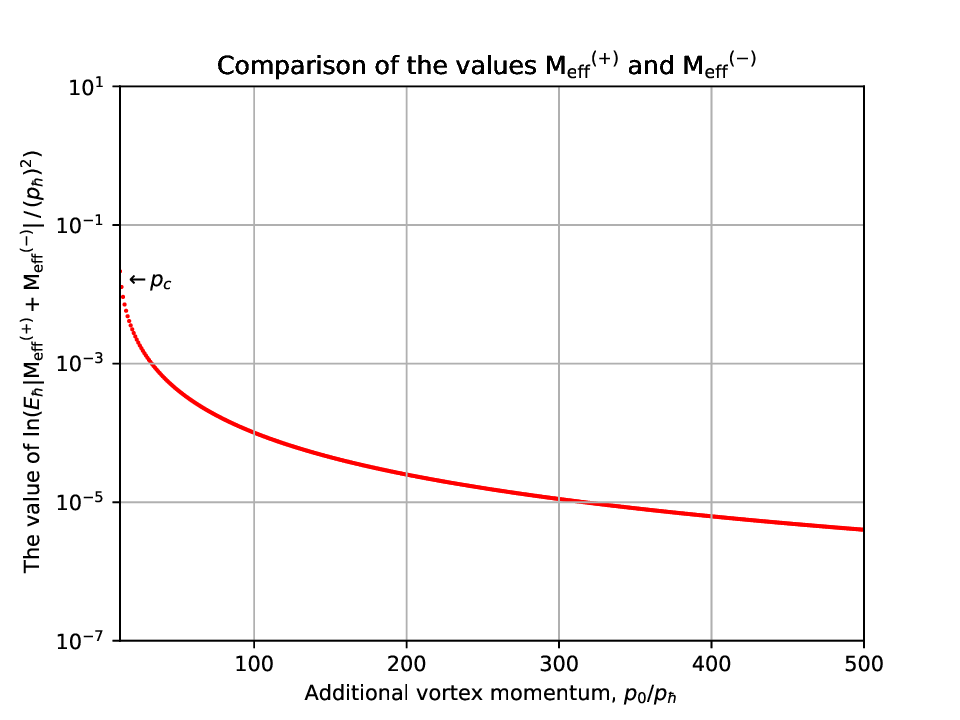}}
{\captionof{figure}{Dependence  $[{\sf M^{(+)}_{eff}}   ~+~   {\sf M^{(-)}_{eff}}]$  on  $p_0$, ~~$\sigma^2_{ph} = 10^{\,-6}$,   ~~$\omega = 10^{\,-9}$.
\label{M_limits}}}
\end{center}


\begin{center}
		 {\includegraphics[width=5.0in]{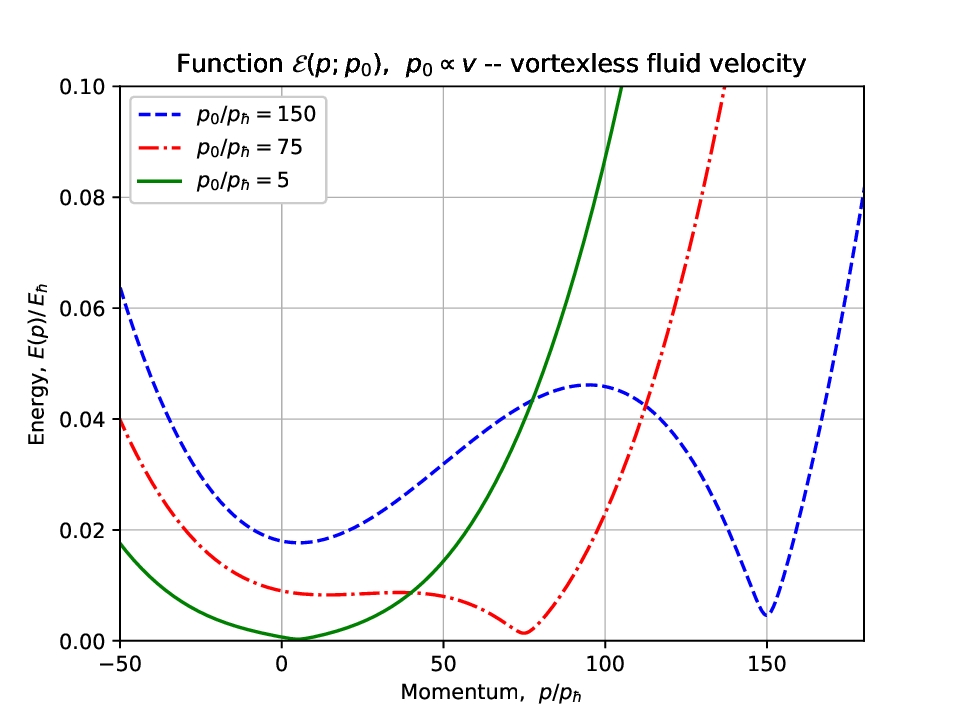}}
{\captionof{figure}{ The transformation of the function $ E = E_{1,1}(p; p_0)$ 
 with respect to the parameter  $p_0$ ~~($\omega = 10^{-3}$).
\label{Ep_dep2}}}
\end{center}


\subsection*{5. Concluding remarks} 

This research explored a separate circular vortex moving through a fluid flow.   We believe that the unique characteristics of the quantum system we are examining might influence the development of a turbulent quantum flow in some way.

{~ The expansion of the considered theory in the many-vortex system has been discussed in previous works by the author. Despite this, we outline here some key points of our approach.
The simplest system  consists of a number of non-interacting circular vortices.
The state space of such a quantum system is as follows:
\[\boldsymbol{H}_K ~=~  \underbrace{\,\boldsymbol{H}_1 \otimes\dots\otimes\boldsymbol{H}_1\,}_{K}\,,\]
where Hilbert spaces $\boldsymbol{H}_1$ are defined as in Eq.(\ref{space_quant}) for each vortex individually.
The full Hamiltonian takes the form
\[   {\widehat H}_v ~=~ \sum_{k=1}^K  \underbrace{I \otimes I \otimes\dots\otimes}_{k-1} {\widehat H}( \omega_k) \underbrace{\otimes I \otimes \dots\otimes I}_{K - k} \,,\]
where 
\[{\widehat H}( \omega_k) ~=~ \sum_{m,n} \int\!\! dp  E_{m,n}({p}; {p}_0) |\,\psi_{{p};\,0,\,m, n}\rangle\langle \psi_{{p};\,0,\,m, n}\,|\,.\]
As regards of the vortex interaction, our approach allows us  to consider the Fock space
\begin{equation}
	\label{Fock_sp}
{\mathfrak H}  ~=~  \bigoplus_{M=0}^{\infty}\boldsymbol{H}_M \, \nonumber  
\end{equation}

To simplify our conclusions, we consider only non-interacting or weakly  interacting vortices in the system.}

 Indeed, let us assume that a fluid volume $V$ contains a sufficiently large number of moving circular vortices ${\mathfrak V}_i$  with different quantum numbers $m_i$ and $n_i$. 
Let the volume $V_i$ be a small neighborhood of a vortex ${\mathfrak V}_i$. For the sake of clarity, let's  suppose that the following conditions hold:
\[{\mathfrak V}_i ~\in~ V_i\,, \qquad  {\mathfrak V}_j ~\not\in~ V_i\,, ~\text{for}~ j \not= i\,, \qquad i = 1,2,\dots, \,.\] 
It is clear that the vortices ${\mathfrak V}_j$ for $j \not= i$, create the certain fluid velocity  ${\boldsymbol{v}}_i$ in the domain ${\mathfrak V}_i$ {~ that can be sufficiently large for large value of $K$.} This velocity isn't linked to the fluid motion of the vortex ${\mathfrak V}_i$ but can be viewed as an external parameter for the system ${\mathfrak V}_i$. 
Of course, this system is more complex than the system considered in this study. Indeed, here  we consider the evolution of quantum vortices with the external parameter $v$ for a certain domain (see Fig.\ref{Pipe_1}) and under special condition 
${\boldsymbol v}  {\boldsymbol p}\nolinebreak \not=\nolinebreak 0$. Employing the same technique, we can explore more complex domains and boundary conditions. 
As it seems in general, the effective mass in the  Eq.(\ref{M_def}) to exhibit a tensorial character.
We suppose that this approach allows us to calculate the critical quantum Reynolds number as follows:
\[ {\sf Re}_q^c ~=~ \inf_{\omega; m,n}{\mathfrak R}_q^c(\omega; m,n)\,.\]
In more detail, the author plans to investigate these  issues in subsequent studies.


\end{document}